\documentclass[%
 reprint,
 amsmath,amssymb,
 aps,
superscriptaddress]{revtex4-2}

\usepackage{graphicx}
\usepackage{dcolumn}
\usepackage{bm}

\usepackage{xcolor}

\begin{document}

\preprint{APS/123-QED}

\title{Measuring the effect of spatial dimension on hydrodynamic turbulence using direct numerical simulation}

\author{Richard D.J.G. Ho}
 \email{richh@uio.no}
 \affiliation{Njord Centre, Department of Physics, University of Oslo, Oslo, 0371, Norway}
 \affiliation{Department of Clinical and Molecular Medicine, Faculty of Medicine and Health Sciences, NTNU, Trondheim, 7491, Norway}
\author{Daniel Clark}
\author{Andres Armua}
\author{Xichao Yang}
\author{Daniel J. Brener}
\author{Arjun Berera}%
 \email{ab@ed.ac.uk}
\affiliation{School of Physics and Astronomy, University of Edinburgh, Edinburgh EH9 3FD, UK
}%

\date{\today}

\begin{abstract}
We perform direct numerical simulation of the incompressible Navier-Stokes equation with forcing at different spatial dimensions and measure turbulent and chaotic properties.
Lyapunov exponents, $\lambda$, decrease with dimension, and $\lambda < 0$ for all simulations in six-dimensions up to $Re = 40$.
These six-dimensional simulations display non-Gaussian statistics and other behavior similar to well developed turbulence despite their lack of chaos.
Further, we find that small scale perturbations do not extend to the largest scales and that this terminal scale between correlation and decorrelation shrinks with dimension.
We theorize that this change is related to the increased rate of vortex stretching.
We find the interplay between turbulent and chaotic properties changes
with increasing dimension.
\end{abstract}

\maketitle



$Introduction.$\textemdash The Navier-Stokes equations (NSEs) in $d$-dimensions are a set of $d$ non-linear partial differential equations which display sensitive dependence on initial conditions: i.e. chaos.
Chaotic properties have emerged as an additional avenue for studying the NSEs, with the advantage of robustness of results at lower Reynolds numbers, $Re$, compared to spectral measures \cite{ho2024chaotic}.
Using direct numerical simulation (DNS), the advantage of using chaotic properties has already been shown in 
analysis of thin-layer turbulence, where the finite time Lyapunov exponent (FTLE) displays a discontinuous jump as the dimension is changed, with a transition between different dynamics \cite{clark2021chaotic}.
Outside DNS, numerical solutions to an eddy damped quasi-normal Markovian
(EDQNM) approximation for the NSEs have found
increased forward energy transfer at higher dimensions,
with the enstrophy production term reaching a maximum near five dimensions \cite{clark2021effect}.
Remarkably, a transition was later found to a non-chaotic regime above $d = 5.88$ \cite{clark2022critical},
with no positive Lyapunov exponent measured even at very high Reynolds numbers.
These results recaptured some of the same trends seen in 4d DNS \cite{berera2020homogeneous}.

Analyzing a system's physical properties whilst changing dimensionality has had a rich history of striking results.
This was perhaps best demonstrated
in the field of critical phenomena with the discovery of
the Wilson-Fischer renormalization group fixed point \cite{wilson1972critical}.
This success in the early 1970s naturally prompted the
application of similar methods with dimension as an analytic parameter in fluid turbulence \cite{kraichnan1959structure,fournier1978infinite}, though this
program has proven to have had limited success.
Difficulty in the theoretical analysis of the NSEs arises due to
the turbulent regime being in the strong coupling limit,
analogous to the problems encountered in quantum chromodynamics
\cite{cheung2020scattering}. 
Even so, Liao developed
an approach, very similar to the Wilson-Fisher theory for deriving
critical phenomena, for the freely decaying isotropic and homogeneous Navier-Stokes turbulence \cite{liao1990some,liao1991kolmogorov}. 
Liao derived to the one-loop level the Kolmogorov exponents for
the energy spectrum of freely-decaying, fully-developed,
near-incompressible turbulence, and found to first order an IR stable fixed-line with upper critical dimension of 6, which has also been found in other renormalization group approaches \cite{verma2024critical}.
The results from EDQNM corroborate Liao’s work suggesting that
interesting behavioral changes occur at six dimensions in the NSEs \cite{clark2022critical}.

Prior to Liao, Nelkin proposed, through a similar analogy to critical phenomena, the existence of a crossover
dimension, above which the Kolmogorov scaling relations in K41 \cite{kolmogorov1962refinement} become exact \cite{nelkin1974turbulence,nelkin1975scaling}. Following Nelkin,
Kraichnan derived that temporal fluctuations of a passive scalar field acting as a tracer for an incompressible
$d$-dimensional turbulent flow were suppressed by increasing dimension \cite{kraichnan1974convection}. Around the same time, these ideas
led to the use of dynamic renormalization group methods by Forster et al. \cite{forster1976long,forster1977large} and DeDomincis and Martin \cite{dedominicis1979energy}.
Further analytical progress in this direction has proved
challenging leaving numerical approaches, in particular
DNS, as the optimal methodology with which to explore
fluid turbulence in higher dimensions \cite{gotoh2007statistical,yamamoto2012local}.
The insight we have added to these earlier studies has been to also utilize chaotic measures in studying the
dimensional dependence of fluid turbulence \cite{ho2024chaotic}. Very different from spectral measures, chaotic measures give a local measure
of the flow and we have found them to be stable from
fairly small box sizes \cite{ho2020fluctuations}.
DNS results for spectral measures and scaling are prohibitively
computational intensive at higher dimensions, whereas
chaotic measures could provide reliable indicators and trends of possible
changes in system behavior already at modest box sizes.

In this Letter, we analyze the properties of the NSEs using DNS from 3 to 6 dimensions.
When using $N^d$ collocation points, computational cost rises faster than $d$, and high $Re$ DNS of homogeneous isotropic turbulence (HIT) are prohibitively expensive.
Since chaotic properties of turbulence have shown promise in analysis of turbulence at lower $Re$, this makes them the most suitable measure to identify changing behavior with dimension.



$Simulations.$\textemdash The NSEs are simulated with a pseudospectral method using code modified from the one described in \cite{yoffe2013investigation,ho2019thesis,berera2020homogeneous}. Similar crosschecks and benchmarks were tested on this code as in these references just stated.
The forcing employed is a delta-correlated in time stochastic force only applied to the low-$k$ (large spatial scale) wavenumbers.

To measure the chaotic properties of the system, we principally use the 
difference field $\textbf{u}_\Delta = \textbf{u}_1 - \textbf{u}_2$, which has associated uncorrelated energy $E_\Delta(k)$, representing the error,
and the finite time Lyapunov exponent (FTLE) \cite{ott2002chaos}, $\lambda$, defined by
\begin{equation}
    \lambda_i(\Delta t) = \frac{1}{\Delta t} \text{ln} \left( \frac{| \textbf{u}_\Delta(\Delta t)|}{\delta_0} \right) \ , \nonumber
\end{equation}
with the difference field allowed to evolve over a time interval $\Delta t$, and $\textbf{u}_1$ rescaled to its original size of $\delta_0$ by 
$\textbf{u}_1 = \textbf{u}_2 + (\textbf{u}_1 - \textbf{u}_2)/\delta_0$.
The FTLE is then found after averaging over many iterations, $i$.

The FTLE is normalized by the timescale $T_{E_0} = E/\varepsilon$, being the ratio of total energy to dissipation.
The Reynolds number, $Re = UL /\nu$, where $\nu$ is the viscosity and
\begin{equation}
    U^2 = \frac{2}{d}\int_0^\infty \text{d}kE(k) \ , \ L=\frac{\Gamma  \left( \frac{d}{2} \right) \sqrt{\pi}}{\Gamma \left( \frac{d+1}{2}\right)U^2} \int_0^\infty \text{d}k \frac{E(k)}{k} \ , \nonumber
\end{equation}
which depend on spatial dimension, $d$.
FTLE were taken after a statistically steady state of $E(k)$ was reached, which occurred much earlier for higher dimensions.
The range of simulation parameters are summarized in Table~\ref{tab:simulations}.

\begin{table}
    \centering
    \begin{tabular}{c|ccc}
        $d$ & $Re$ & $N^d$ & $\lambda T_{E0}$ \\
       \hline
        3 & 5.0 - 1790 & 128 - 1024 & $-$0.19 $\rightarrow$ 16.73 \\
        4 & 22.4 - 184 & 32 - 256 & $-$0.02 $\rightarrow$ 1.90 \\
        5 & 21.0 - 100 & 32 - 96 & $-$0.24 $\rightarrow$ 0.72 \\
        6 & 21.2 - 40.4 & 32 - 48 & $-$0.39 $\rightarrow$ $-$0.16 \\
    \end{tabular}
    \caption{Range of simulation parameters}
    \label{tab:simulations}
\end{table}

\begin{figure}
\includegraphics[width=0.99\linewidth]{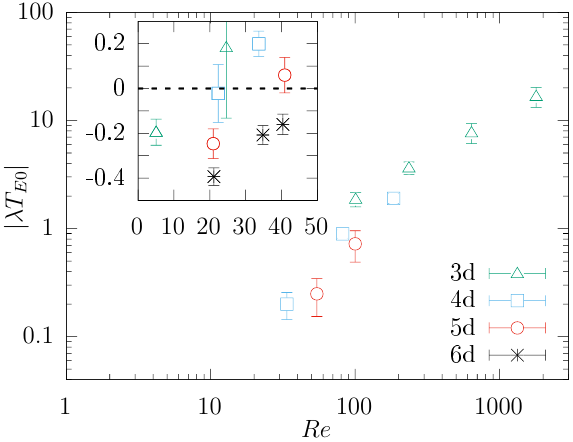}
\caption{\label{fig:lyapvsTE0} Reynolds number dependence of $|\lambda T_{E0}|$, inset shows same data but in linear scale. For six dimensions, the Lyapunov exponent is always negative.}
\end{figure}

$Results.$\textemdash Figure~\ref{fig:lyapvsTE0} shows the dependence of the FTLE with Reynolds number at different dimensions.
The results here are similar to the ones seen in EDQNM simulations \cite{clark2022critical}, where there was an approach to a scaling $\lambda \sim Re^{-\alpha}$ with $\alpha$ independent of dimension,
but with a divergence at low $Re$ indicating a crossover from negative to positive Lyapunov exponent. This crossover point increases with dimension in both these DNS results and the EDQNM results.
The results here also show that, as we increase dimension for a fixed Reynolds number, $\lambda$ decreases, which agrees with EDQNM results.
Previous DNS results in 3d have shown that box sizes have little to no effect on measured Lyapunov exponents in HIT \cite{ho2020fluctuations}, so
we expect that the 6d simulations here are non-chaotic due to the low $Re$, independent of box size.
The $\lambda$ values found here at relatively low $Re$ are similar to those found in EDQNM even at high $Re$, where no dependence on $Re$ was seen. In EDQNM, $\lambda < 0$ was found in 6d even for very high $Re$ and we anticipate that the same may be found in larger DNS simulations, since all trends between the two compare well.

The results above suggest two possibilities, either there is a change at six dimensions where there is no chaos, or that chaos does occur but at much larger $Re$.
The rest of this article will further analyze these simulations to find out to what extent do our low $Re$ 6d simulations indicate turbulent features.

One well known aspect of turbulence is its non-Gaussianity.
From dimension 3 to 5, non-Gaussian velocity derivative skewness has been seen in simulations for decaying turbulence \cite{yamamoto2012local}.
For fully developed turbulence, even at moderate $Re$, the statistics of the Lyapunov exponents has been shown to be non-Gaussian as well in 3d \cite{ho2020fluctuations}.

\begin{figure}
\includegraphics[width=0.99\linewidth]{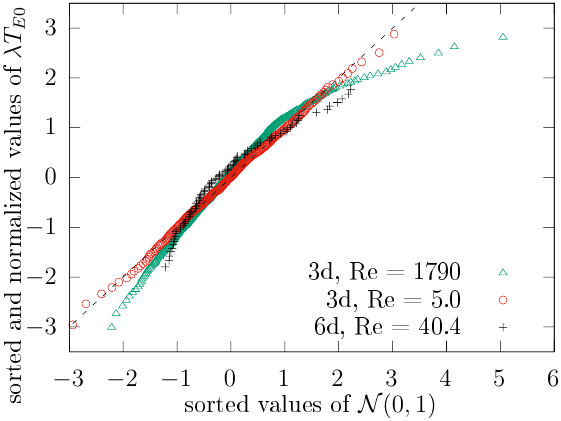}
\caption{\label{fig:qqplot} QQ plot of normalized values. Low $Re$ for 3d is Gaussian (indicated by the straight line dependence), whereas 6d and high $Re$ 3d are not.}
\end{figure}

One tool to measure the deviation of measured data from a target distribution, in this case the normal distribution, is using quantile-quantile (QQ) plots.
All results are sorted in ascending order for the measured data and sample data from the targeted distribution, with a straight line indicating identical distributions.
Figure~\ref{fig:qqplot} shows the QQ plot for the FTLE for three simulations. The FTLE are normalized so that they have zero mean and unit variance.
As can be seen, the values for low $Re$ in 3d show good agreement with a normal distribution.
The highest $Re$ simulation in 6d at $Re = 40.4$ has a curvature more similar to the higher $Re$ 3d simulation at $Re = 1790$, despite that this 6d and low $Re$ 3d simulations have very similar Lyapunov exponents.
This suggests that these simulations both display similar turbulent features, but with the crucial difference that the 6d simulation is non-chaotic.

Further signs that the 6d simulations  at $Re \sim 40$ act like the turbulent high $Re$ 3d simulations and unlike non-turbulent low $Re$ 3d simulations can be found by analyzing the error spectrum, $E_\Delta(k)$.
This measures the energy spectrum associated with the difference field of the two realizations, $u_\Delta = u_1 - u_2$.

\begin{figure}
\includegraphics[width=0.95\linewidth]{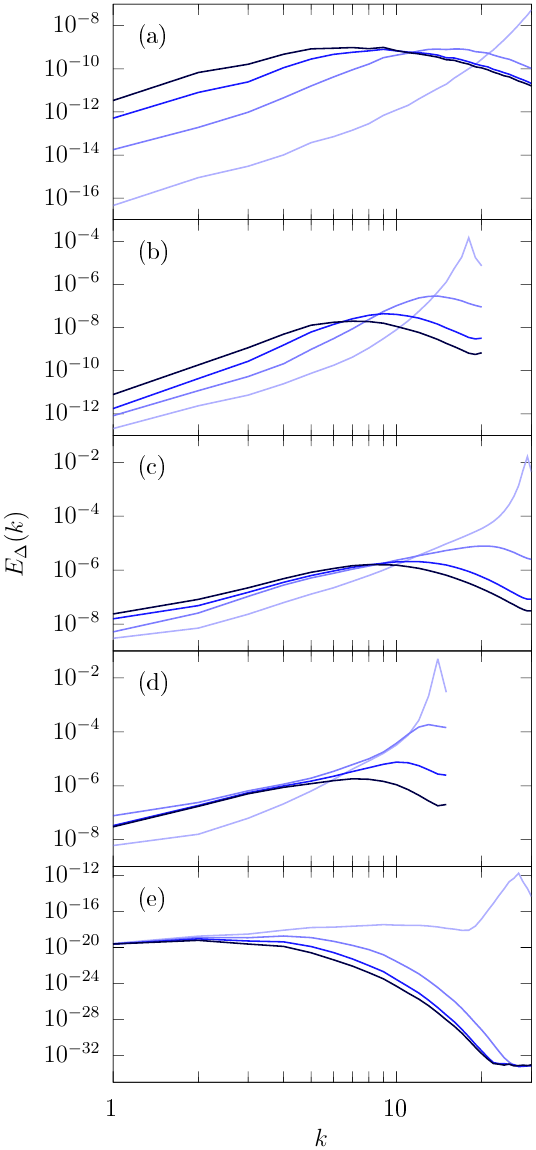}
\caption{\label{fig:edk} Evolution of difference between two realisations with time, from early (light blue) to later (darker blue) times, showing different behavior with higher dimension: a) 3d $Re$ = 101, b) 4d $Re$ = 33.7, c) 5d $Re$ = 100, d) 6d $Re$ = 40.4, e) 3d $Re$ = 3.5.}
\end{figure}

In Figure~\ref{fig:edk} we show the effect of making a perturbation at high-$k$ (small spatial scale), which in 3-5d tends to feed differences at lower $k$ (larger scales) for sufficient $Re$, i.e. where $\lambda > 0$.
As can be seen, although the 6d $(d)$ and 3d low $Re$ $(e)$ simulations have similar negative $\lambda$, $(d)$ does not look like $(e)$ but instead is qualitatively similar to the simulations performed at lower dimensions with sufficient $Re$ for $\lambda > 0$ $(a-c)$.
After a perturbation is made,
two processes occur: there is the transfer of error from high-$k$ to low-$k$; and also a generation of more error at high-$k$.
These are balanced by the fact that the forcing of both fields is identical, which has a strong organizing tendency.
As we increase dimension, as shown in the figure, the amount of inverse transfer of error is reduced.
In 6d, the error at low-$k$ arises due to the very large initial perturbation spreading from higher to lower wavenumber at a transfer rate that is larger than the organizing effect can counteract; the error generated subsequently at high-$k$ is self sustaining but unable to transfer upwards as it is counteracted by the stronger forward cascade.
In low $Re$ 3d simulations, the dissipation is high enough to simply destroy the error at high-$k$, which is not the case for 6d.
Thus we see, in regimes where neither are chaotic, 3d behaves fundamentally differently than 6d.

\begin{figure}
\includegraphics[width=0.99\linewidth]{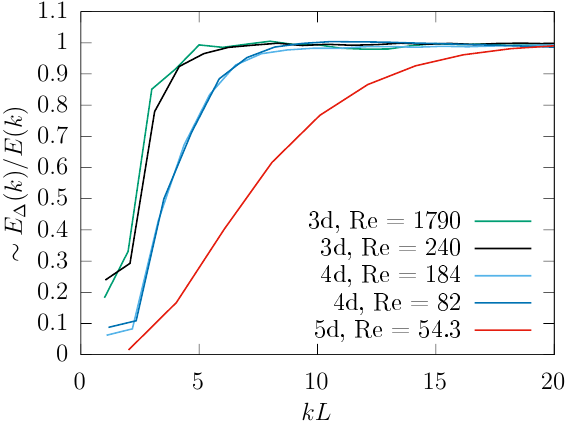}
\caption{\label{fig:edksteady} Steady state of difference at late times, showing reduced steady state of low wavenumber with higher dimension.}
\end{figure}

EDQNM models showed a transition to a non-chaotic regime as we approach $d=6$ \cite{clark2022critical}, and our DNS results confirm many of the findings using those models,
this includes that an initial small scale error does not saturate at the lowest wavenumber, as shown in Figure~\ref{fig:edksteady}, which plots the ratio of $E_\Delta(k)$ to $E(k)$ when $E_\Delta(t)$ is statistically stationary in time.
We see that the error ceases to grow at sizes slightly smaller than the largest in the system, well below the largest theoretical value (being $E(k)$).
From the figure, it can be seen that the effect is dependent on dimension and not on the Reynolds number.
These results are different to those found previously using DNS in 4d \cite{berera2020homogeneous}, but can be explained by the different forcing functions, as discussed in \cite{ge2023production,ge2025eulerian}. Previous simulations used negative damping, which amplifies error right to the forcing scale.
The use of an identical forcing function reveals that there is significant suppression of error growth at larger scales with increasing dimension regardless of $Re$.

\begin{figure}
\includegraphics[width=0.99\linewidth]{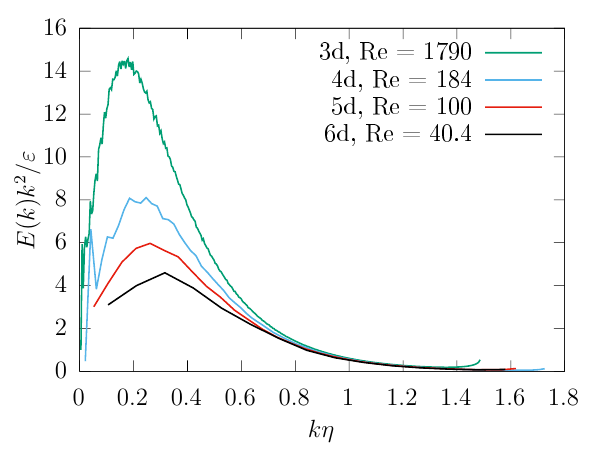}
\caption{\label{fig:enstrophy} Enstrophy spectrum, with the peak shifting to higher wavenumber with increasing spatial dimension.}
\end{figure}

The washing out of correlation at larger length scales may be caused by the interaction of enstrophy production and self-strain vortex stretching,
where enstrophy conservation is found when removing the vortex stretching term in three dimensions \cite{bos2021three}. A deeper discussion on how the two relate to chaos is found in \cite{ge2023production,ge2025interscale}.
There is a competition between the generation of error moving from high-$k$ to low-$k$
and the sweeping out of error by the energy cascade.
In 3d, the balance is achieved at a low value of $kL$,
but with increasing dimension it is shifted
higher as the competition favors the sweeping.
Figure~\ref{fig:enstrophy} shows the enstrophy spectrum at different dimensions, and it is clearly seen that the peak goes to higher wavenumbers with increasing dimension. This higher wavenumber represents an increase in the efficiency of the enstrophy production, and has been seen in EDQNM models \cite{clark2021effect,clark2022critical,clark2022effect}.
The more efficient cascade will sweep out the error before it can spread to larger scales.

One can understand the growth of uncorrelated energy as a sum of three terms, the internal production of uncertainty,
the dissipation of uncertainty,
and its external input rate via the difference between the forcing of the two fields \cite{ge2023production}, where the last one is kept at zero for these simulations.
It has been shown that local compressions increase the internal production of uncertainty, whereas local stretching reduces uncertainty \cite{ge2023production,ge2025interscale}.
The more efficient enstrophy cascade represents an increased local stretching, and is sufficient to suppress the growth of uncorrelated energy.

$Discussion.$\textemdash In 3d at $Re$ $\approx 40$, it is well understood the system exhibits turbulent
features, such as in QQ plots and evolution of difference.
In 6d, our simulations find similar features.
However, in sharp contrast, the 6d simulations at this $Re$ are non-chaotic whereas in 3d they are chaotic.
Much larger box sizes than those used here in 6d would be needed to examine more
conventional measures of turbulent behavior and K41 features, such as the energy
spectrum, as well as for
making useful visualizations.  
This is a very big computational requirement.  
The forced simulations done in this Letter have a largest box size in 6d of $48^6$,
which in terms of collocation points is equivalent to a 3d simulation of $2304^3$
and in 3d would be sufficient for usable statistics.
Even increasing the box size to $100^6$, which is still relatively small,
would push this to be among the larger DNS simulations performed.
This paper has identified measures that can be utilized at
smaller sizes to find the first telltale hints of changing behavior as dimension increases.

We have found that at least up to $Re \approx 40$, 
6d turbulence is non-chaotic.
Due to the small box sizes, we cannot say what the implications are for the K41 theory in higher dimensions.
However, we find in DNS that with increasing dimension: $\lambda$ decreases; and the largest decorrelated size in the system shrinks.
This is consistent with work from EDQNM closures, where a transition to non-chaotic behavior was seen at six dimensions \cite{clark2022critical}.
This effect can be understood using the framework of an uncertainty cascade in turbulence, which has recently been studied in three dimensions \cite{ge2025interscale,vela2025uncertainty}, but may change with dimension.
Previous analysis has shown that the production of uncertainty is associated with local compression events, whereas local stretching reduced the uncertainty \cite{clark2021effect,ge2023production,ge2024thesis}.
With increasing dimension, the forward cascade is enhanced because energetic triads become more dominated by orthogonal wave vectors \cite{fournier1978infinite}.
All of these suggest that in higher dimensions the flow of information becomes unidirectional, with flow lines tending to move without much interaction, with the small scales unable to influence the large.

Hydrodynamics is the quintessential field theory, being described in field theoretic terms in a similar era to electromagnetism, and certainly before quantum theory.
Our collective experience with (quantum) field theory has taught
us the importance for theoretical understanding of studying
higher dimensions beyond the physical regime.
This Letter is the first DNS study of Navier-Stokes
turbulence up to six spatial dimensions.
It examined both the turbulent and chaotic properties of
the flow, finding a deviation in the interplay
between the two as the system dimension increases,
at least to the extent of our limited simulations.

In turbulence, we have a notion that the smallest scales cannot affect the largest scales. This is true in a statistical sense. 
At sufficiently high dimension,
this may be to be true in an absolute sense as well.
HIT is often considered as occurring at scales where the turbulence ``forgets'' about the larger scales at which it is generated. Here we show that, as dimension increases, the larger scales ``ignore'' what is happening at the smaller scales altogether.
In the language of chaos theory, a butterfly flapping its
wings being able to affect a hurricane at the other side of
the world is true in three dimensions, but at higher dimensions
this work suggests that effect diminishes, 
though larger simulations would be needed
to confirm the trends found here.

$Acknowledgements.$\textemdash This work used resources from ARCHER II \cite{archer} via the Director's Time budget.

$Data \ availability.$\textemdash The data that support the findings of this article are openly available \cite{ho_2025_17622323}.

\bibliography{ndimturb}

\begin{thebibliography}{34}%
\makeatletter
\providecommand \@ifxundefined [1]{%
 \@ifx{#1\undefined}
}%
\providecommand \@ifnum [1]{%
 \ifnum #1\expandafter \@firstoftwo
 \else \expandafter \@secondoftwo
 \fi
}%
\providecommand \@ifx [1]{%
 \ifx #1\expandafter \@firstoftwo
 \else \expandafter \@secondoftwo
 \fi
}%
\providecommand \natexlab [1]{#1}%
\providecommand \enquote  [1]{``#1''}%
\providecommand \bibnamefont  [1]{#1}%
\providecommand \bibfnamefont [1]{#1}%
\providecommand \citenamefont [1]{#1}%
\providecommand \href@noop [0]{\@secondoftwo}%
\providecommand \href [0]{\begingroup \@sanitize@url \@href}%
\providecommand \@href[1]{\@@startlink{#1}\@@href}%
\providecommand \@@href[1]{\endgroup#1\@@endlink}%
\providecommand \@sanitize@url [0]{\catcode `\\12\catcode `\$12\catcode `\&12\catcode `\#12\catcode `\^12\catcode `\_12\catcode `\%12\relax}%
\providecommand \@@startlink[1]{}%
\providecommand \@@endlink[0]{}%
\providecommand \url  [0]{\begingroup\@sanitize@url \@url }%
\providecommand \@url [1]{\endgroup\@href {#1}{\urlprefix }}%
\providecommand \urlprefix  [0]{URL }%
\providecommand \Eprint [0]{\href }%
\providecommand \doibase [0]{https://doi.org/}%
\providecommand \selectlanguage [0]{\@gobble}%
\providecommand \bibinfo  [0]{\@secondoftwo}%
\providecommand \bibfield  [0]{\@secondoftwo}%
\providecommand \translation [1]{[#1]}%
\providecommand \BibitemOpen [0]{}%
\providecommand \bibitemStop [0]{}%
\providecommand \bibitemNoStop [0]{.\EOS\space}%
\providecommand \EOS [0]{\spacefactor3000\relax}%
\providecommand \BibitemShut  [1]{\csname bibitem#1\endcsname}%
\let\auto@bib@innerbib\@empty
\bibitem [{\citenamefont {Ho}\ \emph {et~al.}(2024)\citenamefont {Ho}, \citenamefont {Clark},\ and\ \citenamefont {Berera}}]{ho2024chaotic}%
  \BibitemOpen
  \bibfield  {author} {\bibinfo {author} {\bibfnamefont {R.~D.}\ \bibnamefont {Ho}}, \bibinfo {author} {\bibfnamefont {D.}~\bibnamefont {Clark}},\ and\ \bibinfo {author} {\bibfnamefont {A.}~\bibnamefont {Berera}},\ }\bibfield  {title} {\bibinfo {title} {Chaotic measures as an alternative to spectral measures for analysing turbulent flow},\ }\href@noop {} {\bibfield  {journal} {\bibinfo  {journal} {Atmosphere}\ }\textbf {\bibinfo {volume} {15}},\ \bibinfo {pages} {1053} (\bibinfo {year} {2024})}\BibitemShut {NoStop}%
\bibitem [{\citenamefont {Clark}\ \emph {et~al.}(2021{\natexlab{a}})\citenamefont {Clark}, \citenamefont {Armua}, \citenamefont {Freeman}, \citenamefont {Brener},\ and\ \citenamefont {Berera}}]{clark2021chaotic}%
  \BibitemOpen
  \bibfield  {author} {\bibinfo {author} {\bibfnamefont {D.}~\bibnamefont {Clark}}, \bibinfo {author} {\bibfnamefont {A.}~\bibnamefont {Armua}}, \bibinfo {author} {\bibfnamefont {C.}~\bibnamefont {Freeman}}, \bibinfo {author} {\bibfnamefont {D.~J.}\ \bibnamefont {Brener}},\ and\ \bibinfo {author} {\bibfnamefont {A.}~\bibnamefont {Berera}},\ }\bibfield  {title} {\bibinfo {title} {Chaotic measure of the transition between two-and three-dimensional turbulence},\ }\href@noop {} {\bibfield  {journal} {\bibinfo  {journal} {Physical Review Fluids}\ }\textbf {\bibinfo {volume} {6}},\ \bibinfo {pages} {054612} (\bibinfo {year} {2021}{\natexlab{a}})}\BibitemShut {NoStop}%
\bibitem [{\citenamefont {Clark}\ \emph {et~al.}(2021{\natexlab{b}})\citenamefont {Clark}, \citenamefont {Ho},\ and\ \citenamefont {Berera}}]{clark2021effect}%
  \BibitemOpen
  \bibfield  {author} {\bibinfo {author} {\bibfnamefont {D.}~\bibnamefont {Clark}}, \bibinfo {author} {\bibfnamefont {R.~D.}\ \bibnamefont {Ho}},\ and\ \bibinfo {author} {\bibfnamefont {A.}~\bibnamefont {Berera}},\ }\bibfield  {title} {\bibinfo {title} {Effect of spatial dimension on a model of fluid turbulence},\ }\href@noop {} {\bibfield  {journal} {\bibinfo  {journal} {Journal of Fluid Mechanics}\ }\textbf {\bibinfo {volume} {912}},\ \bibinfo {pages} {A40} (\bibinfo {year} {2021}{\natexlab{b}})}\BibitemShut {NoStop}%
\bibitem [{\citenamefont {Clark}\ \emph {et~al.}(2022)\citenamefont {Clark}, \citenamefont {Armua}, \citenamefont {Ho},\ and\ \citenamefont {Berera}}]{clark2022critical}%
  \BibitemOpen
  \bibfield  {author} {\bibinfo {author} {\bibfnamefont {D.}~\bibnamefont {Clark}}, \bibinfo {author} {\bibfnamefont {A.}~\bibnamefont {Armua}}, \bibinfo {author} {\bibfnamefont {R.~D.}\ \bibnamefont {Ho}},\ and\ \bibinfo {author} {\bibfnamefont {A.}~\bibnamefont {Berera}},\ }\bibfield  {title} {\bibinfo {title} {Critical transition to a non-chaotic regime in isotropic turbulence},\ }\href@noop {} {\bibfield  {journal} {\bibinfo  {journal} {Journal of Fluid Mechanics}\ }\textbf {\bibinfo {volume} {930}},\ \bibinfo {pages} {A17} (\bibinfo {year} {2022})}\BibitemShut {NoStop}%
\bibitem [{\citenamefont {Berera}\ \emph {et~al.}(2020)\citenamefont {Berera}, \citenamefont {Ho},\ and\ \citenamefont {Clark}}]{berera2020homogeneous}%
  \BibitemOpen
  \bibfield  {author} {\bibinfo {author} {\bibfnamefont {A.}~\bibnamefont {Berera}}, \bibinfo {author} {\bibfnamefont {R.~D.}\ \bibnamefont {Ho}},\ and\ \bibinfo {author} {\bibfnamefont {D.}~\bibnamefont {Clark}},\ }\bibfield  {title} {\bibinfo {title} {Homogeneous isotropic turbulence in four spatial dimensions},\ }\href@noop {} {\bibfield  {journal} {\bibinfo  {journal} {Physics of Fluids}\ }\textbf {\bibinfo {volume} {32}} (\bibinfo {year} {2020})}\BibitemShut {NoStop}%
\bibitem [{\citenamefont {Wilson}\ and\ \citenamefont {Fisher}(1972)}]{wilson1972critical}%
  \BibitemOpen
  \bibfield  {author} {\bibinfo {author} {\bibfnamefont {K.~G.}\ \bibnamefont {Wilson}}\ and\ \bibinfo {author} {\bibfnamefont {M.~E.}\ \bibnamefont {Fisher}},\ }\bibfield  {title} {\bibinfo {title} {Critical exponents in 3.99 dimensions},\ }\href@noop {} {\bibfield  {journal} {\bibinfo  {journal} {Physical Review Letters}\ }\textbf {\bibinfo {volume} {28}},\ \bibinfo {pages} {240} (\bibinfo {year} {1972})}\BibitemShut {NoStop}%
\bibitem [{\citenamefont {Kraichnan}(1959)}]{kraichnan1959structure}%
  \BibitemOpen
  \bibfield  {author} {\bibinfo {author} {\bibfnamefont {R.~H.}\ \bibnamefont {Kraichnan}},\ }\bibfield  {title} {\bibinfo {title} {The structure of isotropic turbulence at very high reynolds numbers},\ }\href@noop {} {\bibfield  {journal} {\bibinfo  {journal} {Journal of Fluid Mechanics}\ }\textbf {\bibinfo {volume} {5}},\ \bibinfo {pages} {497} (\bibinfo {year} {1959})}\BibitemShut {NoStop}%
\bibitem [{\citenamefont {Fournier}\ \emph {et~al.}(1978)\citenamefont {Fournier}, \citenamefont {Frisch},\ and\ \citenamefont {Rose}}]{fournier1978infinite}%
  \BibitemOpen
  \bibfield  {author} {\bibinfo {author} {\bibfnamefont {J.-D.}\ \bibnamefont {Fournier}}, \bibinfo {author} {\bibfnamefont {U.}~\bibnamefont {Frisch}},\ and\ \bibinfo {author} {\bibfnamefont {H.}~\bibnamefont {Rose}},\ }\bibfield  {title} {\bibinfo {title} {Infinite-dimensional turbulence},\ }\href@noop {} {\bibfield  {journal} {\bibinfo  {journal} {Journal of Physics A: Mathematical and General}\ }\textbf {\bibinfo {volume} {11}},\ \bibinfo {pages} {187} (\bibinfo {year} {1978})}\BibitemShut {NoStop}%
\bibitem [{\citenamefont {Cheung}\ and\ \citenamefont {Mangan}(2020)}]{cheung2020scattering}%
  \BibitemOpen
  \bibfield  {author} {\bibinfo {author} {\bibfnamefont {C.}~\bibnamefont {Cheung}}\ and\ \bibinfo {author} {\bibfnamefont {J.}~\bibnamefont {Mangan}},\ }\bibfield  {title} {\bibinfo {title} {Scattering amplitudes and the navier-stokes equation},\ }\href@noop {} {\bibfield  {journal} {\bibinfo  {journal} {arXiv preprint arXiv:2010.15970}\ } (\bibinfo {year} {2020})}\BibitemShut {NoStop}%
\bibitem [{\citenamefont {Liao}(1990)}]{liao1990some}%
  \BibitemOpen
  \bibfield  {author} {\bibinfo {author} {\bibfnamefont {W.}~\bibnamefont {Liao}},\ }\bibfield  {title} {\bibinfo {title} {Some ideas on the freely decaying navier-stokes turbulence},\ }\href@noop {} {\bibfield  {journal} {\bibinfo  {journal} {Journal of Physics A: Mathematical and General}\ }\textbf {\bibinfo {volume} {23}},\ \bibinfo {pages} {L159} (\bibinfo {year} {1990})}\BibitemShut {NoStop}%
\bibitem [{\citenamefont {Liao}(1991)}]{liao1991kolmogorov}%
  \BibitemOpen
  \bibfield  {author} {\bibinfo {author} {\bibfnamefont {W.}~\bibnamefont {Liao}},\ }\bibfield  {title} {\bibinfo {title} {Kolmogorov exponents for near-incompressible turbulence from perturbative quantum field theory},\ }\href@noop {} {\bibfield  {journal} {\bibinfo  {journal} {Journal of statistical physics}\ }\textbf {\bibinfo {volume} {65}},\ \bibinfo {pages} {1} (\bibinfo {year} {1991})}\BibitemShut {NoStop}%
\bibitem [{\citenamefont {Verma}(2024)}]{verma2024critical}%
  \BibitemOpen
  \bibfield  {author} {\bibinfo {author} {\bibfnamefont {M.~K.}\ \bibnamefont {Verma}},\ }\bibfield  {title} {\bibinfo {title} {Critical dimension for hydrodynamic turbulence},\ }\href@noop {} {\bibfield  {journal} {\bibinfo  {journal} {Physical Review E}\ }\textbf {\bibinfo {volume} {110}},\ \bibinfo {pages} {035102} (\bibinfo {year} {2024})}\BibitemShut {NoStop}%
\bibitem [{\citenamefont {Kolmogorov}(1962)}]{kolmogorov1962refinement}%
  \BibitemOpen
  \bibfield  {author} {\bibinfo {author} {\bibfnamefont {A.~N.}\ \bibnamefont {Kolmogorov}},\ }\bibfield  {title} {\bibinfo {title} {A refinement of previous hypotheses concerning the local structure of turbulence in a viscous incompressible fluid at high reynolds number},\ }\href@noop {} {\bibfield  {journal} {\bibinfo  {journal} {Journal of Fluid Mechanics}\ }\textbf {\bibinfo {volume} {13}},\ \bibinfo {pages} {82} (\bibinfo {year} {1962})}\BibitemShut {NoStop}%
\bibitem [{\citenamefont {Nelkin}(1974)}]{nelkin1974turbulence}%
  \BibitemOpen
  \bibfield  {author} {\bibinfo {author} {\bibfnamefont {M.}~\bibnamefont {Nelkin}},\ }\bibfield  {title} {\bibinfo {title} {Turbulence, critical fluctuations, and intermittency},\ }\href@noop {} {\bibfield  {journal} {\bibinfo  {journal} {Physical Review A}\ }\textbf {\bibinfo {volume} {9}},\ \bibinfo {pages} {388} (\bibinfo {year} {1974})}\BibitemShut {NoStop}%
\bibitem [{\citenamefont {Nelkin}(1975)}]{nelkin1975scaling}%
  \BibitemOpen
  \bibfield  {author} {\bibinfo {author} {\bibfnamefont {M.}~\bibnamefont {Nelkin}},\ }\bibfield  {title} {\bibinfo {title} {Scaling theory of hydrodynamic turbulence},\ }\href@noop {} {\bibfield  {journal} {\bibinfo  {journal} {Physical Review A}\ }\textbf {\bibinfo {volume} {11}},\ \bibinfo {pages} {1737} (\bibinfo {year} {1975})}\BibitemShut {NoStop}%
\bibitem [{\citenamefont {Kraichnan}(1974)}]{kraichnan1974convection}%
  \BibitemOpen
  \bibfield  {author} {\bibinfo {author} {\bibfnamefont {R.~H.}\ \bibnamefont {Kraichnan}},\ }\bibfield  {title} {\bibinfo {title} {Convection of a passive scalar by a quasi-uniform random straining field},\ }\href@noop {} {\bibfield  {journal} {\bibinfo  {journal} {Journal of fluid mechanics}\ }\textbf {\bibinfo {volume} {64}},\ \bibinfo {pages} {737} (\bibinfo {year} {1974})}\BibitemShut {NoStop}%
\bibitem [{\citenamefont {Forster}\ \emph {et~al.}(1976)\citenamefont {Forster}, \citenamefont {Nelson},\ and\ \citenamefont {Stephen}}]{forster1976long}%
  \BibitemOpen
  \bibfield  {author} {\bibinfo {author} {\bibfnamefont {D.}~\bibnamefont {Forster}}, \bibinfo {author} {\bibfnamefont {D.~R.}\ \bibnamefont {Nelson}},\ and\ \bibinfo {author} {\bibfnamefont {M.~J.}\ \bibnamefont {Stephen}},\ }\bibfield  {title} {\bibinfo {title} {Long-time tails and the large-eddy behavior of a randomly stirred fluid},\ }\href@noop {} {\bibfield  {journal} {\bibinfo  {journal} {Physical Review Letters}\ }\textbf {\bibinfo {volume} {36}},\ \bibinfo {pages} {867} (\bibinfo {year} {1976})}\BibitemShut {NoStop}%
\bibitem [{\citenamefont {Forster}\ \emph {et~al.}(1977)\citenamefont {Forster}, \citenamefont {Nelson},\ and\ \citenamefont {Stephen}}]{forster1977large}%
  \BibitemOpen
  \bibfield  {author} {\bibinfo {author} {\bibfnamefont {D.}~\bibnamefont {Forster}}, \bibinfo {author} {\bibfnamefont {D.~R.}\ \bibnamefont {Nelson}},\ and\ \bibinfo {author} {\bibfnamefont {M.~J.}\ \bibnamefont {Stephen}},\ }\bibfield  {title} {\bibinfo {title} {Large-distance and long-time properties of a randomly stirred fluid},\ }\href@noop {} {\bibfield  {journal} {\bibinfo  {journal} {Physical Review A}\ }\textbf {\bibinfo {volume} {16}},\ \bibinfo {pages} {732} (\bibinfo {year} {1977})}\BibitemShut {NoStop}%
\bibitem [{\citenamefont {DeDominicis}\ and\ \citenamefont {Martin}(1979)}]{dedominicis1979energy}%
  \BibitemOpen
  \bibfield  {author} {\bibinfo {author} {\bibfnamefont {C.}~\bibnamefont {DeDominicis}}\ and\ \bibinfo {author} {\bibfnamefont {P.}~\bibnamefont {Martin}},\ }\bibfield  {title} {\bibinfo {title} {Energy spectra of certain randomly-stirred fluids},\ }\href@noop {} {\bibfield  {journal} {\bibinfo  {journal} {Physical Review A}\ }\textbf {\bibinfo {volume} {19}},\ \bibinfo {pages} {419} (\bibinfo {year} {1979})}\BibitemShut {NoStop}%
\bibitem [{\citenamefont {Gotoh}\ \emph {et~al.}(2007)\citenamefont {Gotoh}, \citenamefont {Watanabe}, \citenamefont {Shiga}, \citenamefont {Nakano},\ and\ \citenamefont {Suzuki}}]{gotoh2007statistical}%
  \BibitemOpen
  \bibfield  {author} {\bibinfo {author} {\bibfnamefont {T.}~\bibnamefont {Gotoh}}, \bibinfo {author} {\bibfnamefont {Y.}~\bibnamefont {Watanabe}}, \bibinfo {author} {\bibfnamefont {Y.}~\bibnamefont {Shiga}}, \bibinfo {author} {\bibfnamefont {T.}~\bibnamefont {Nakano}},\ and\ \bibinfo {author} {\bibfnamefont {E.}~\bibnamefont {Suzuki}},\ }\bibfield  {title} {\bibinfo {title} {Statistical properties of four-dimensional turbulence},\ }\href@noop {} {\bibfield  {journal} {\bibinfo  {journal} {Physical Review E—Statistical, Nonlinear, and Soft Matter Physics}\ }\textbf {\bibinfo {volume} {75}},\ \bibinfo {pages} {016310} (\bibinfo {year} {2007})}\BibitemShut {NoStop}%
\bibitem [{\citenamefont {Yamamoto}\ \emph {et~al.}(2012)\citenamefont {Yamamoto}, \citenamefont {Shimizu}, \citenamefont {Inoshita}, \citenamefont {Nakano},\ and\ \citenamefont {Gotoh}}]{yamamoto2012local}%
  \BibitemOpen
  \bibfield  {author} {\bibinfo {author} {\bibfnamefont {T.}~\bibnamefont {Yamamoto}}, \bibinfo {author} {\bibfnamefont {H.}~\bibnamefont {Shimizu}}, \bibinfo {author} {\bibfnamefont {T.}~\bibnamefont {Inoshita}}, \bibinfo {author} {\bibfnamefont {T.}~\bibnamefont {Nakano}},\ and\ \bibinfo {author} {\bibfnamefont {T.}~\bibnamefont {Gotoh}},\ }\bibfield  {title} {\bibinfo {title} {Local flow structure of turbulence in three, four, and five dimensions},\ }\href@noop {} {\bibfield  {journal} {\bibinfo  {journal} {Physical Review E—Statistical, Nonlinear, and Soft Matter Physics}\ }\textbf {\bibinfo {volume} {86}},\ \bibinfo {pages} {046320} (\bibinfo {year} {2012})}\BibitemShut {NoStop}%
\bibitem [{\citenamefont {Ho}\ \emph {et~al.}(2020)\citenamefont {Ho}, \citenamefont {Armua},\ and\ \citenamefont {Berera}}]{ho2020fluctuations}%
  \BibitemOpen
  \bibfield  {author} {\bibinfo {author} {\bibfnamefont {R.~D.}\ \bibnamefont {Ho}}, \bibinfo {author} {\bibfnamefont {A.}~\bibnamefont {Armua}},\ and\ \bibinfo {author} {\bibfnamefont {A.}~\bibnamefont {Berera}},\ }\bibfield  {title} {\bibinfo {title} {Fluctuations of lyapunov exponents in homogeneous and isotropic turbulence},\ }\href@noop {} {\bibfield  {journal} {\bibinfo  {journal} {Physical Review Fluids}\ }\textbf {\bibinfo {volume} {5}},\ \bibinfo {pages} {024602} (\bibinfo {year} {2020})}\BibitemShut {NoStop}%
\bibitem [{\citenamefont {Yoffe}(2013)}]{yoffe2013investigation}%
  \BibitemOpen
  \bibfield  {author} {\bibinfo {author} {\bibfnamefont {S.~R.}\ \bibnamefont {Yoffe}},\ }\bibfield  {title} {\bibinfo {title} {Investigation of the transfer and dissipation of energy in isotropic turbulence},\ }\href@noop {} {\bibfield  {journal} {\bibinfo  {journal} {arXiv preprint arXiv:1306.3408}\ } (\bibinfo {year} {2013})}\BibitemShut {NoStop}%
\bibitem [{\citenamefont {Ho}(2019)}]{ho2019thesis}%
  \BibitemOpen
  \bibfield  {author} {\bibinfo {author} {\bibfnamefont {R.~D.}\ \bibnamefont {Ho}},\ }\bibfield  {title} {\bibinfo {title} {Effects of macroscopic variables on turbulent evolution},\ }\href@noop {} {\bibfield  {journal} {\bibinfo  {journal} {PhD Thesis}\ } (\bibinfo {year} {2019})}\BibitemShut {NoStop}%
\bibitem [{\citenamefont {Ott}(2002)}]{ott2002chaos}%
  \BibitemOpen
  \bibfield  {author} {\bibinfo {author} {\bibfnamefont {E.}~\bibnamefont {Ott}},\ }\href@noop {} {\emph {\bibinfo {title} {Chaos in dynamical systems}}}\ (\bibinfo  {publisher} {Cambridge university press},\ \bibinfo {year} {2002})\BibitemShut {NoStop}%
\bibitem [{\citenamefont {Ge}\ \emph {et~al.}(2023)\citenamefont {Ge}, \citenamefont {Rolland},\ and\ \citenamefont {Vassilicos}}]{ge2023production}%
  \BibitemOpen
  \bibfield  {author} {\bibinfo {author} {\bibfnamefont {J.}~\bibnamefont {Ge}}, \bibinfo {author} {\bibfnamefont {J.}~\bibnamefont {Rolland}},\ and\ \bibinfo {author} {\bibfnamefont {J.~C.}\ \bibnamefont {Vassilicos}},\ }\bibfield  {title} {\bibinfo {title} {The production of uncertainty in three-dimensional navier--stokes turbulence},\ }\href@noop {} {\bibfield  {journal} {\bibinfo  {journal} {Journal of Fluid Mechanics}\ }\textbf {\bibinfo {volume} {977}},\ \bibinfo {pages} {A17} (\bibinfo {year} {2023})}\BibitemShut {NoStop}%
\bibitem [{\citenamefont {Ge}\ \emph {et~al.}(2025{\natexlab{a}})\citenamefont {Ge}, \citenamefont {Rolland},\ and\ \citenamefont {Vassilicos}}]{ge2025eulerian}%
  \BibitemOpen
  \bibfield  {author} {\bibinfo {author} {\bibfnamefont {J.}~\bibnamefont {Ge}}, \bibinfo {author} {\bibfnamefont {J.}~\bibnamefont {Rolland}},\ and\ \bibinfo {author} {\bibfnamefont {J.~C.}\ \bibnamefont {Vassilicos}},\ }\bibfield  {title} {\bibinfo {title} {Eulerian-lagrangian scaling of the lyapunov exponent in homogeneous turbulence},\ }\href@noop {} {\bibfield  {journal} {\bibinfo  {journal} {Physical Review Fluids}\ }\textbf {\bibinfo {volume} {10}},\ \bibinfo {pages} {L072601} (\bibinfo {year} {2025}{\natexlab{a}})}\BibitemShut {NoStop}%
\bibitem [{\citenamefont {Bos}(2021)}]{bos2021three}%
  \BibitemOpen
  \bibfield  {author} {\bibinfo {author} {\bibfnamefont {W.~J.}\ \bibnamefont {Bos}},\ }\bibfield  {title} {\bibinfo {title} {Three-dimensional turbulence without vortex stretching},\ }\href@noop {} {\bibfield  {journal} {\bibinfo  {journal} {Journal of Fluid Mechanics}\ }\textbf {\bibinfo {volume} {915}},\ \bibinfo {pages} {A121} (\bibinfo {year} {2021})}\BibitemShut {NoStop}%
\bibitem [{\citenamefont {Ge}\ \emph {et~al.}(2025{\natexlab{b}})\citenamefont {Ge}, \citenamefont {Rolland},\ and\ \citenamefont {Vassilicos}}]{ge2025interscale}%
  \BibitemOpen
  \bibfield  {author} {\bibinfo {author} {\bibfnamefont {J.}~\bibnamefont {Ge}}, \bibinfo {author} {\bibfnamefont {J.}~\bibnamefont {Rolland}},\ and\ \bibinfo {author} {\bibfnamefont {J.~C.}\ \bibnamefont {Vassilicos}},\ }\bibfield  {title} {\bibinfo {title} {The interscale behaviour of uncertainty in three-dimensional navier--stokes turbulence},\ }\href@noop {} {\bibfield  {journal} {\bibinfo  {journal} {Journal of Fluid Mechanics}\ }\textbf {\bibinfo {volume} {1017}},\ \bibinfo {pages} {A29} (\bibinfo {year} {2025}{\natexlab{b}})}\BibitemShut {NoStop}%
\bibitem [{\citenamefont {Clark}(2022)}]{clark2022effect}%
  \BibitemOpen
  \bibfield  {author} {\bibinfo {author} {\bibfnamefont {D.}~\bibnamefont {Clark}},\ }\bibfield  {title} {\bibinfo {title} {Effect of spatial dimensionality on the chaotic properties of turbulent flow},\ }\href@noop {} {\bibfield  {journal} {\bibinfo  {journal} {PhD Thesis}\ } (\bibinfo {year} {2022})}\BibitemShut {NoStop}%
\bibitem [{\citenamefont {Vela-Mart{\'\i}n}(2025)}]{vela2025uncertainty}%
  \BibitemOpen
  \bibfield  {author} {\bibinfo {author} {\bibfnamefont {A.}~\bibnamefont {Vela-Mart{\'\i}n}},\ }\bibfield  {title} {\bibinfo {title} {The uncertainty cascade in isotropic turbulence},\ }\href@noop {} {\bibfield  {journal} {\bibinfo  {journal} {Journal of Fluid Mechanics}\ }\textbf {\bibinfo {volume} {1021}},\ \bibinfo {pages} {A8} (\bibinfo {year} {2025})}\BibitemShut {NoStop}%
\bibitem [{\citenamefont {Ge}(2024)}]{ge2024thesis}%
  \BibitemOpen
  \bibfield  {author} {\bibinfo {author} {\bibfnamefont {J.}~\bibnamefont {Ge}},\ }\bibfield  {title} {\bibinfo {title} {Evolution of uncertainty in three-dimensional navier-stokes turbulence},\ }\href@noop {} {\bibfield  {journal} {\bibinfo  {journal} {PhD Thesis}\ } (\bibinfo {year} {2024})}\BibitemShut {NoStop}%
\bibitem [{arc()}]{archer}%
  \BibitemOpen
  \href {https://www.archer2.ac.uk/} {https://www.archer2.ac.uk/}\BibitemShut {NoStop}%
\bibitem [{\citenamefont {Ho}\ \emph {et~al.}(2025)\citenamefont {Ho}, \citenamefont {Clark}, \citenamefont {Armua}, \citenamefont {Yang}, \citenamefont {Brener},\ and\ \citenamefont {Berera}}]{ho_2025_17622323}%
  \BibitemOpen
  \bibfield  {author} {\bibinfo {author} {\bibfnamefont {R.}~\bibnamefont {Ho}}, \bibinfo {author} {\bibfnamefont {D.}~\bibnamefont {Clark}}, \bibinfo {author} {\bibfnamefont {A.}~\bibnamefont {Armua}}, \bibinfo {author} {\bibfnamefont {X.}~\bibnamefont {Yang}}, \bibinfo {author} {\bibfnamefont {D.~J.}\ \bibnamefont {Brener}},\ and\ \bibinfo {author} {\bibfnamefont {A.}~\bibnamefont {Berera}},\ }\bibfield  {title} {\bibinfo {title} {Dataset for "measuring the effect of spatial dimension on hydrodynamic turbulence using direct numerical simulation"},\ }\href {https://doi.org/10.5281/zenodo.17622323} {10.5281/zenodo.17622323} (\bibinfo {year} {2025})\BibitemShut {NoStop}%
\end{thebibliography}%

\end{document}